\documentclass[useAMS,usenatbib]{mn2e}
\usepackage{amssymb}
\usepackage{graphicx}
\usepackage{mathrsfs}
\usepackage{bm}
\title[The Entrainment-Limited Evolution of FR\,II sources]{The Entrainment-Limited Evolution of FR\,II Sources: Maximum Sizes and A Possible Connection to FR\,Is} \author[Y. Wang et al.]{Y. Wang$^{1}$\thanks{E-mail:
ywang@mpifr.de} C. Knigge$^{2}$ J. H. Croston$^{2}$ and G. Pavlovski$^{2}$\\$^{1}$Max-Planck-Institute for Radioastronomy, Auf dem H\"ugel 69, Bonn 53121, Germany\\$^{2}$School of Physics and Astronomy, University of Southampton, University Road, Southampton SO17 1BJ, United Kingdom}

\begin{document}

\pagerange{\pageref{firstpage}--\pageref{lastpage}} \pubyear{2002}

\maketitle

\label{firstpage}

% partially modified

\begin{abstract}
We construct a simple theoretical model to investigate how entrainment gradually erodes high-speed FR\,II jets. This process is described by embedding a mixing-layer model developed originally to describe FR\,I objects in a self-similar model for the lobe structure of classical FR\,II sources. Following the classical FR\,II models, we assume that the lobe is dominated by the particles injected from the central jet. The entrainment produces a boundary shear layer which acts at the interface between the dense central jet and the less denser surrounding lobe, and the associated erosion of the jet places interesting limits on the maximum size of FR\,II sources. The model shows that this limit depends mainly on the initial bulk velocity of the relativistic jet triggered. The bulk velocities of FR\,IIs suggested by our model are in good agreement with that obtained from direct pc-scale observations on ordinary radio galaxies and quasars. Finally, we discuss how FR\,IIs may evolve into FR\,Is upon reaching their maximum, entrainment-limited sizes.

\end{abstract}


\section{Introduction}
\label{introduction}
Radio galaxies appear to come in two fundamentally different types, as encapsulated in the Fanaroff and Riley classification scheme \citep{fr74}. FR\,I sources have bright cores and edge-darkened lobes, while FR\,II sources are edge-brightened with hotspots at the end. These morphological differences suggest that the interactions between the radio jets and their environments are very different in the two classes, and that their evolutionary tracks may also be quite distinct.

FR\,II sources are powerful radio sources with fairly homogeneous morphologies. They contain highly relativistic jets extending from the central AGN to very bright hotspots surrounded by low surface brightness lobes. Dynamical models for FR\,II sources are quite successful, indicating that the lobes expand in a self-similar way \citep[][hereafter KA97]{falle91,ka97}. Based on this, a range of radio emission models have been developed, and these models allow FR\,II sources to be tracked through the power-linear size (P-D) diagram \citep{kda97,brw99, mk02}.

FR\,I sources are much more common than FR\,IIs \citep{parma02}, but they are also more complex and have only one common feature: no hotspot at the outer end of the jet. About half of the FR\,I sources show a \textit{fat double} morphology similar to FR\,II lobes, while the rest inflate turbulent lobes after passing through a so-called brightening point, with plumes or tails at the end \citep{ol89, ow91, parma02}. Modeling FR\,I sources is difficult, as it is hard to describe all types of FR\,I sources with a single model. \citet{bick94} tried to model the FR\,I sources by relativistic conservation laws, and \citet{lb02a} studied 3C\,31 in detail. \citet[][hereafter W09]{wang09} adopted the mixing-layer structure from \citet[][hereafter CR91]{cr91} and built an analytical model which could explain the observational behavior of typical tailed FR\,I sources.

Generally speaking, FR\,IIs are more powerful than FR\,Is, with a transition radio luminosity around $P_{178\textrm{MHz}}\sim10^{25}$W\,Hz$^{-1}$\,sr$^{-1}$. A transition luminosity also exists in the optical band \citep{ol94}. The value of the transition luminosity is not precise, as it also depends on the properties of the host galaxies \citep{lo96}. The origin of the \textit{FR\,I/II dichotomy} has been discussed extensively in the literature. Studies on compact steep-spectrum sources (CSS) suggested that these objects are typically young and may generally evolve into large-scale radio sources \citep{fanti95}. Among these CSS sources,  FR\,Is and FR\,IIs may have different progenitors due to different powers and environments \citep{alexander00, kunert05}. However, the transition from FR\,IIs to FR\,Is can also possibly take place at a later stage of the jet evolution under certain circumstance \citep{falle91,bicknell95,kb07}.

The instabilities of jets have been studied by a number of numerical simulations, and they find that the jet instability evolution and the large-scale jet morphology are mainly determined by the jet Lorentz factor \citep{perucho04b, perucho05}, and the ambient density profile \citep{rossi08, meliani08}. However, these numerical simulations more concentrated on the earlier stage of the jet evolutions, when the lobe structures around the central relativistic jets have not been well developed and differ from that at the later jet evolution stage. Meanwhile, the initial set of the simulations (e.g. the radial resolution) may also affect the simulation results \citep{perucho04b}. Therefore, analytical models describing the evolutions of jet stabilities are desired for studying the jets at the late stage of their evolutions.

%consider the interactions between the relativistic jets and their environments. This may only work well at the beginning of the jet evolution, and the longest numerical simulations only simulate several $10^{6}$\,yr or $\sim10\%$ of the jet lifetime. For well developed FR\,IIs with larger ages, it may not be applicable since the relativistic jets are surrounded by low density lobes instead of ambient medium. Models including the lobe component are desired if we want to investigate the stabilities of a well developed FR\,II jet. 

Taking the idea of the relativistic mixing layer model developed by W09, we can investigate more precisely how the central jets of FR\,II sources are eroded by the entrainment due to the interactions with their surrounding lobes. More specifically, we describe the entrainment process by embedding the W09 mixing-layer model for FR\,I jets in a simple, self-similar model of an FR\,II radio lobe, and monitor how the central jet is gradually eroded by the growing turbulent shear layer at the interface between the jet and the lobe. The goal of this paper is to study if the entrainment can play an important role on determining the maximum size of an FR\,II jet at the late stage of its evolution. Our basic model for a central jet and shear layer embedded in an FR\,II radio lobe is developed in Section \ref{model_development}. The maximum size of FR\,II jet is calculated and discussed in Section \ref{maximum_size}. We also argue that the resulting \textit{dead} FR\,IIs will ultimately re-emerge as FR\,I radio sources. A brief sketch of this transition process is provided in Section \ref{model_transition}. Finally, in Section \ref{conclusion}, we summarise our conclusions and outline future work suggested by our model.

\section{Model development}
\label{model_development}

\begin{figure}
\includegraphics[width=0.5\textwidth]{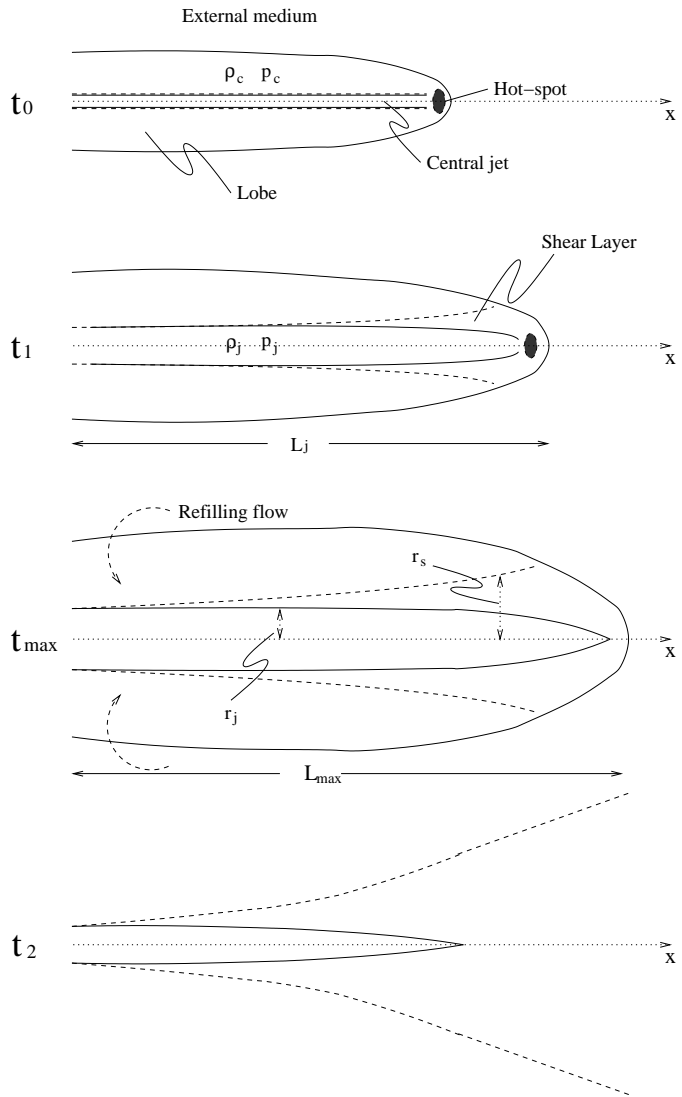}
\caption{Sketch of the evolution of a radio outflow. At $t_{0}$, the young outflow is showing a FR\,II morphology. At $t_{1}$, the outflow is still in FR\,II phase while the shear layer has already grown. At $t_{2}$, the hotspot vanish and the outflow will transfer into FR\,I stage after this age.}
\label{cartoon}
\end{figure}

We describe FR\,II objects by embedding a highly relativistic central jet inside a surrounding radio-emitting lobe. Although the lobe density is thought to be very low, we assume that a turbulent shear layer may nevertheless form at the jet-lobe interface. This shear layer will entrain and mix material from both regions. This entrainment will gradually erode the central jet. More specifically, once all of the highly relativistic material in the central jet has been mixed up with the lobe material in the shear layer, the central jet is completely destroyed. Previous work suggests that the hotspot is a very compact, high-pressure region that gives rise to strong radio emission \citep{scheck02}. The highly relativistic central jet may play an important role in energising the hotspot as there is large amount of energy injected into a small area. Meanwhile, the bulk velocity of the material in the shear layer is not as fast as that in the central jet. Although the shear layer is still supersonic and can form weak shocks and working surfaces, it is not powerful or concentrated enough to support a hotspot. Thus, we assume that as the central jet is gradually eroded, the hotspot weakens. After the central jet is totally destroyed at a certain stage of FR\,II evolution, the hotspot vanishes at the same time. This is in agreement with observations which indicate that the hotspot luminosity decreases with the linear size of the FR\,II source \citep{pm03}. When the hotspot vanishes, the object will cease to be a \textit{proper} FR\,II and will most likely resemble a lobed FR\,I. A sketch of this process is described in Figure \ref{cartoon}. 

Most analytical FR\,II models suggest the lobe is formed by the particles injected from the central jet through the hotspot \citep{falle91}. Although numerical simulations show that there is efficient mixing between the lobe and the shocked ambient medium \citep[e.g.][]{scheck02}, their simulations are not long enough and only represent a relatively early stage of the lobes. X-ray inverse Compton measurements of FR\,II radio lobes show that the strength of the magnetic field in the lobe is close to the equipartition value, which suggests that the FR\,II lobes do not contain an energetically dominant proton population \citep{ks05, croston05}. Meanwhile, the lobe internal pressures are in good agreement with the environment pressures, which suggests that there is no need for substantial mixing to provide the required pressures \citep{Belsole07}. These are the evidence showing that the interactions between the FR\,II lobes and their environments on large scales are not significant. Therefore, as our model here is based on previous analytical models and we are only considering FR\,IIs at a late stage of their evolution, we neglect the mixing and assume all the lobe material is from the central relativistic jet. As the lobe occupies a much bigger space, the density in the lobe is much lower than the density of central relativistic jet.

The AGN active time is thought to be around a few $10^{8}$\,yr, and the maximum size of FR\,II objects is observed to be a few Mpc. In order to decide whether the interaction between the jet and the lobe can ever be a significant factor in the evolution of an FR\,II, we therefore need to consider if entrainment could conceivably destroy the central jet on this time and/or length scale. The mixing-layer model for FR\,I objects developed by W09 discussed the interaction between a laminar jet and its environment. It also predicts the position where the laminar jet disappears. We can therefore apply the same model to the FR\,II case and study the interaction between a jet and its lobe in the relativistic limit. This allows us to place interesting limits on the maximum sizes of FR\,II sources due to entrainment. In this section, we will first outline the analytical FR\,II lobe model and the W09 mixing-layer model, and then present the results with typical values for the parameter.

\subsection{The self-similar model for FR\,II lobes}
KA97 and \citet{kda97} have established a successful model for FR\,II radio sources describing their dynamics and evolutions. In this section, we summarise the important features of the model.

KA97 follow the basic dynamical picture proposed by \citet{scheuer74} and \citet{falle91}, assuming that the laminar jets will end in strong shocks where the electrons are accelerated. The electrons pass through the shocks and subsequently inflate a lobe with a uniformly distributed pressure. The jet is in pressure-equilibrium with the lobe, and KA97 showed that the lobe then expands in a self-similar way.

The evolution of the lobe size is determined by a balance of the ram pressure of the lobe material and that of the medium surrounding the host galaxy, which is pushed aside by the jet. The density distribution outside the core radius, $a_{0}$, is approximated by a power-law, $\rho(x)=\rho_{0}a_{0}^{\alpha}x^{-\alpha}$, where $x$ is the radial distance from the central AGN and $\rho_{0}$ is the density in the core radius, $a_{0}$. KA97 suggested that, for typical radio galaxies, $\rho_{0}=7.2\times10^{-22}$\,kgm$^{-3}$ at $a_{0}=2$\,kpc. These values may vary for different sources, but we will later show that the precise numbers here are not important in our model. The exponent $0<\alpha\le2$ is constrained by both theories and observations. X-ray observations find that the exponents for most clusters are close to 1.5 \citep{vikhlinin06, croston08}, so we adopt $\alpha=1.5$ for the moment and will discuss the effects of adopting different values later in Section 3.2.1.

Having set the density profile above, we can express the length of the lobe by:
\begin{equation}
L_{j}=c_{1}(Q_{0}t^{3}/\Lambda)^{\frac{1}{5-\alpha}},
\end{equation}
where $Q_{0}$ is the jet power, $\Lambda=\rho_{0}a_{0}^{\alpha}$, $t$ is the jet age and $c_{1}$ is a constant given by equation (25) in KA97. The pressure of the lobe also evolves with the jet age and can be written as:
\begin{equation}
p_{c}=\frac{18c_{1}^{2-\alpha}}{(\Gamma_{x}+1)(5-\alpha)^{2}4R_{T}^{2}}\Lambda^{\frac{3}{5-\alpha}}Q_{0}^{\frac{2-\alpha}{5-\alpha}}t^{\frac{-4-\alpha}{5-\alpha}}.
\end{equation}
$\Gamma_{x}$ is the adiabatic index of the external medium, which is set to be 5/3 here. $R_{T}$ is the axial ratio, which is normally distributed between 1.3 and 6, with an average value of 2 \citep{lw84}. For simplicity, we adopt this value initially and will discuss the effect of adopting different values in Section 3.2.2.

As the jet grows in a self-similar way, we can express the volume of the lobe by $V=\pi L_{j}^{3}/(4R_{T}^{2})$. The particles injected into the jet are believed to be highly relativistic, so the rest mass injected into the lobe is given by $m_{0}=Q_{0}t/(c^{2}\gamma_{j})$, where $\gamma_{j}$ is the Lorentz factor of the particles injected. With the expression of $V$ and $m_{0}$, the density in the lobe is given by:
\begin{equation}
\rho_{c}=\frac{m_{0}}{V}=\frac{4R_{T}^{2}}{\pi c^{2}\gamma_{j} c_{1}^{3}}\Lambda^{\frac{3}{5-\alpha}}Q_{0}^{\frac{2-\alpha}{5-\alpha}}t^{\frac{-4-\alpha}{5-\alpha}}.
\end{equation}
We refer the reader to KA97 for a detailed derivation/explanation of the equations above.

\subsection{Entrainment and the mixing-layer model}

W09 model FR\,I sources with a mixing-layer structure in which a laminar jet interacts with its environment by forming a turbulent mixing layer at the interface between the two regions. This growing shear layer continuously entrains and mixes material from the jet and its environment, until finally the laminar core has been completely eroded and disappears. The structure of the different layers is then determined by using relativistic fluid mechanics and applying the relativistic conservation laws of mass, momentum and energy.

In this paper, we borrow this basic picture to estimate under what conditions the central jet of an FR\,II object may disappear. We assume that an FR\,II object evolves following KA97 model. Its central jet is therefore embedded inside the lobe and is presumably subject to entrainment from the lobe. The interaction of the jet with the lobe, and its subsequent evolution, are described by the relativistic mixing-layer model from W09. It is important to note that in the case of FR\,II type objects, the central jets are not in direct contact with the environment, in contrast to the model for FR\,Is presented in W09.  Unlike the external medium, the properties of the material inside the lobe, e.g. the pressure and the density, are assumed to have uniform distributions. Thus we adopt the \textit{constant environment} case of the W09 FR\, I model with $p_{e}=p_{c}$ and $\rho=\rho_{c}$, where $p_{c}$ and $\rho_{c}$ are given by Equations (2) and (3) respectively.

For this simplified case, the three relativistic conservation laws are re-written as:

{\setlength\arraycolsep{1pt}
\begin{eqnarray}
\frac{\mathscr{R}_{j}\Gamma_{j}}{\Gamma_{j}-1}&&\gamma_{j}\beta_{j}(r_{0}^{2}-r_{j}^{2}(x))=\nonumber\\ &&\frac{\mathscr{R}_{s}(x)\Gamma_{s}}{\Gamma_{s}-1}\gamma_{s}\beta_{s}(r_{s}^{2}(x)-r_{j}^{2}(x))-F(x),\\
\frac{(\mathscr{R}_{j}+1)\Gamma_{j}}{\Gamma_{j}-1}&&\gamma_{j}^{2}\beta_{j}^{2}(r_{0}^{2}-r_{j}^{2}(x))=\nonumber\\ &&\frac{(\mathscr{R}_{s}(x)+1)\Gamma_{s}}{\Gamma_{s}-1}\gamma_{s}^{2}\beta_{s}^{2}(r_{s}^{2}(x)-r_{j}^{2}(x)),\\
\frac{(\mathscr{R}_{j}+1)\Gamma_{j}}{\Gamma_{j}-1}&&\gamma_{j}^{2}\beta_{j}(r_{0}^{2}-r_{j}^{2}(x))=\nonumber\\ &&\frac{(\mathscr{R}_{s}(x)+1)\Gamma_{s}}{\Gamma_{s}-1}\gamma_{s}^{2}\beta_{s}(r_{s}^{2}(x)-r_{j}^{2}(x))-F(x).
\end{eqnarray}
}

Based on the equations above, the radius of the central jet, $r_{j}$ could be expressed as a function of the distance away from the central AGN, $x$:

\begin{equation}
r_{j}(x)^{2}=r_{0}^{2}-\frac{F(x)(\Gamma_{j}-1)}{\Gamma_{j}\gamma_{j}^2\beta_{j}(\frac{\beta_{j}}{\beta_{s}}-1)(\mathscr{R}_{j}+1)},
\end{equation}
where $\Gamma_{j}=\Gamma_{s}=4/3$ are the adiabatic indices of the material inside the central jet and the shear layer. $r_{0}$ is the initial radius of the jet at the brightening point, which we assume is a constant equal to 100\,pc throughout the life time of the jet.

$\mathscr{R}_{j}$ is defined as the ratio between rest mass energy and non-relativistic enthalpy for jet material. In principle, as jet pressure decreases with jet age, the value of $\mathscr{R}_{j}$ should increase. However, this value may vary for different sources and it is hard to estimate from the observation. W09 obtained $\mathscr{R}_{j}=13.4$ by applying this entrainment model to 3C\,31, which is an old FR\,I source. Considering we are discussing FR\,II sources at the late stage of their evolution, we assume a common value of $\mathscr{R}_{j}=10$ for simplicity in our calculations.

$F(x)=cg(x)/[\pi p(x)]$ is defined in W09, where $g(x)=\int_{S}\rho\bm{v_{\rm ent}}\bm{\cdot n}dS$ is the mass entrainment function ($\bm{n}$ is the normal direction of the unit surface $dS$). Taking Equations (2) and (3), we find that $F(x)$ is is given by:
\begin{equation}
F(x)=\frac{8R_{T}^{4}(\Gamma_{x}+1)(5-\alpha)^{2}}{\pi^{2}c\gamma_{j}c_{1}^{5-\alpha}}\int_{S}\bm{v_{\rm ent}}\bm{\cdot n}dS.
\end{equation}
As discussed in W09, entrainment is mainly due to turbulent motions, so the entrainment velocity $\bm{v_{ent}}$ is closely related to the sound speed, $C_{c}$ in the lobe. This is a constant throughout the jet lifetime, as the lobe is undergoing adiabatic expansion. We set $v_{ent}=\eta C_{c}$, and $\eta$ is the entrainment efficiency which is set to be 0.5 here. KA97 and CR91 defined and used an entrainment efficiency in much the same way and argued for an upper limit of $\eta<0.26$ in their non-relativistic mixing layer model. Our model is built under different conditions, and we choose a default value of $\eta=0.5$. We discuss this issue in more detail in Section 3.2.3. As $\int_{S}dS$ is also a function of $r_{j}(x)$, we find that $r_{j}(x)$ is just a function of $\gamma_{j}$ and could be solved numerically. This interestingly shows that $L_{max}$, the maximum distance that an FR\,II object can reach, where $r_{j}(L_{max})=0$, only depends on the Lorentz factor, $\gamma_{j}$.

$\beta=v/c$ and $\gamma=(1-\beta^{2})^{-0.5}$ are measures of the bulk velocity. $\beta_{j}$ and $\beta_{s}$ refer to the velocities in the central jet and the shear layer defined by W09 respectively. The analysis of some typical FR\,I sources indicate that bulk velocities are $\beta\approx$ 0.8 -- 0.9 where the jets first brighten abruptly and decelerate rapidly to speeds of $\beta \approx$ 0.1 -- 0.4 where recollimation takes place \citep[e.g.][]{lb02a, cl04, canvin05, lcbh06}. Both the analytical models and numerical simulations suggest that the jet velocity has a transverse structure. However, it has a restricted range and does not evolve significantly with $x$ \citep{lb02b}, so it is reasonable to use a quasi-one-dimensional analysis and adopt velocity values averaged across the jet cross-section in each region (W09). In this paper, we adopt $\beta_{s}=0.3$ as a typical value. The value of $\beta_{j}$, which is a key value deciding the maximum length of FR\,II object, will be discussed in the following section.

\section{The maximum size of FR\,II jets}
\label{maximum_size}

\subsection{Results}
\label{result}

\begin{figure}
\includegraphics[width=0.5\textwidth]{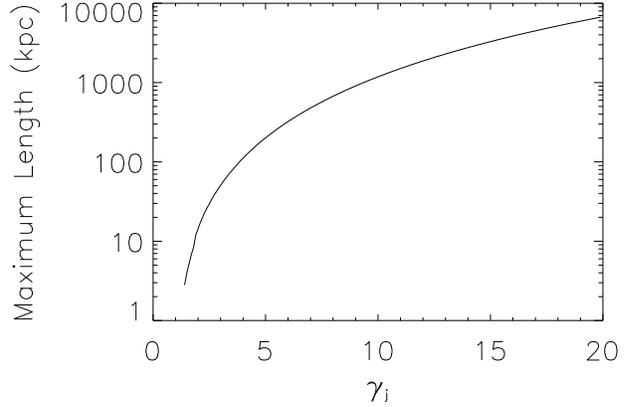}
\caption{The maximum length of FR\,II jet as a function of $\beta_{j}$, with $\alpha=1.5$ and $R_{T}=2$.}
\label{gamma_l}
\end{figure}

The model discussed in the last section contains various parameters and most of them have been fixed based on previous observations or theoretical work. We first concentrate on the maximum length that an FR\,II jet can reach with given jet bulk velocity. Figure \ref{gamma_l} shows the distribution of $L_{max}$ as a function of $\gamma_{j}$. The diagram suggests that, if the particles in the central jet are slow, $L_{max}$ is small, and the FR\,II structure will be easily destroyed by entrainment at an early stage of its evolution. For example, an FR\,II with $\gamma_{j}=2$ ($\beta_{j}\sim0.87$) can survive only to a maximum length of $\sim5$ kpc. As $\beta_{j}$ increases, the jet can survive longer and the maximum jet length increases. An FR\,II with $\gamma_{j}=15$ ($\beta_{j}\sim0.998$) can reach as far as 3000\,kpc. This upper limit is sufficient to cover almost all the FR\,IIs currently observed.

\begin{figure}
\includegraphics[width=0.5\textwidth]{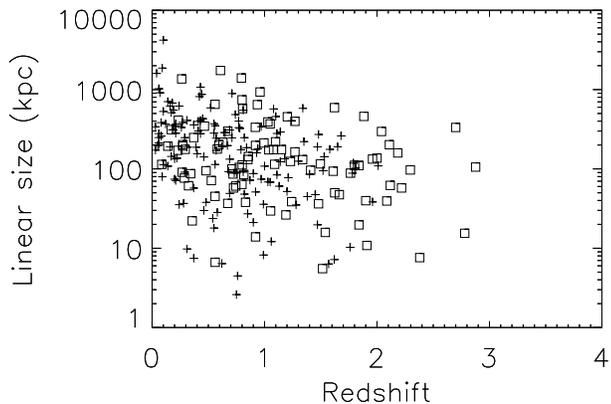}
\caption{The distributions of 3CRR and 7CRS sources on the linear size-redshift plane. The plus signs are 3CRR sources and the square signs are 7CRS sources.} \label{l_z}
\end{figure}

Relativistic beaming has been observed for many radio sources, and it is widely accepted that powerful FR\,IIs have a very high bulk velocity. Direct measurement of jet bulk velocity is difficult, but several attempts have been made. \citet{hough02} observed the parsec-scale regions of 25 quasars (with FR\,II morphology) in the 3CRR sample and estimated their bulk velocities to fall in the range of $\gamma_{j}\approx5-10$. X-ray observations done by \citet{sambruna04} require a $\gamma_{j}\sim10$. \citet{hardcastle06} applied beamed inverse-Compton model to a sample of X-ray jets and shows the required bulk velocity on parsec scale could be as high as $\gamma_{j}\ge15$. \citet{begelman08} obtained similar results implying $\gamma_{j}\geq2$ and possibly as high as $\sim50$. \citet{jorstad08} studied a sample of radio loud galaxies and stated that the mean $\gamma_{j}$ is about 5 for radio galaxies, about 13 for BL Lacs, and about 20 for quasars. However, all these estimations of the bulk velocities are quite rough and model dependent. For example, although \citet{hough02} measured the apparent velocities for individual sources, their orientation angles were not well constrained. \citet{hardcastle06} used a fixed angle for all their sources. Considering that the apparent velocities of all sources are superluminal, the value of orientation angle plays an important role on calculating the real bulk velocity. The uncertainty of the orientation angles together with the uncertainty of apparent velocity measurements imply huge uncertainties on the calculated real jet velocities and real jet sizes. We therefore cannot convincingly test our model by comparing predictions with observations of individual systems. Nevertheless, we can check for the overall consistency in a statistical sense here. Observational samples of FR\,IIs (e.g. 3CRR, 6CE and 7CRS) show that most objects have sizes between 10\,kpc and 1000\,kpc. These size scales refer to a $\gamma_{j}\gtrsim2-10$ based on our model. The largest object in the 3CRR sample has a length of $\sim4000$kpc, which refers to $\gamma_{j}\gtrsim17$. All these values of Lorentz factor required can be fulfilled by the observational results described above.

\citet{mh09} analysed a complete sample of FR\,IIs and suggested that although the jet bulk speed on parsec scales can be as high as $\gamma>10$, it might be much smaller on kpc scales. Their fitted model gave a Lorentz factor around $1.18-1.49$, corresponding to a speed of $0.53c-0.74c$. This is also consistent with some earlier work implying that the large-scale jets only have moderately relativistic bulk speeds \citep[e.g.][]{wa97,hardcastle99,al04}. However, this deceleration phenomenon can be naturally explained by our model here: the speeds observed at kpc scale may refer to the region containing both the central jet and the mixing layer. As the bulk velocity in the layer can be $0.3c$ or even lower, it is possible for us to observe a moderately relativistic velocity when the boundary layer dominates the jet on kpc scale.

We can also check the maximum age of an FR\,II jet. With the observational constrained Lorentz factors, our model predicts that most FR\,IIs can have a maximum size between 10 kpc and a few 1000 kpc. \citet{oc98} observed young and powerful Compact Symmetric Objects, (which are believed to be the progenitors of radio galaxies,) and obtained a hotspot advance speed of $\sim0.03c-0.3c$ at a scale of $\sim100$\,pc. Deceleration of the head may happen as a slower speed of $\sim0.02c$ was obtained for kpc scale jet Cygnus A \citep{cb96}. Therefore, we can estimate a maximum age range from several $10^{5}$\,yr up to $10^{8}$\,yr, which indicates that an FR\,II jet can be destroyed anytime between these time scales, but it can hardly survive beyond $10^{8}$ yr. This range of maximum age agrees well with other analytical work, e.g. \citet{bird08}, who found an average jet age of $1.2\times10^{7}$ yr.

The most interesting conclusion from our model is that, as most parameters of the model are restricted in a small range by the observations, the maximum length of an FR\,II object is mainly determined by the bulk velocity of the central jet. Jets with a similar bulk velocity will have a similar $L_{max}$. Here we would like to investigate two complete samples, 3CRR and 7CRS, to see if their source distributions support this idea. It is hard to obtain accurate jet properties for individuals, but as complete samples, we can assume the overall distributions of the jet properties (e.g. jet size, age and environment density) represent the average value at each redshift. \citet{wk08} find there is a strong relation between the redshift and FR\,II environment density ($\Lambda\propto(1+z)^{5.8}$). If $L_{max}$ is affected by the environment, we should observe a strong relation between the average jet size and the redshift. Figure \ref{l_z} shows that the overall jet size only slightly decreases by a factor of $\sim3$ from local universe to $z=2$. This small decrease may be more due to the selection effect: The sources are fainter when they get older and larger, so at high redshift, the sources easily fall below the flux limit before they grow to a fairly large size. The observations are not in contrast with our model's prediction, but better obtained jet properties or larger complete samples are surely desired in the future work to provide a better support for our model.

Our model also indicates that $L_{max}$ does not depend directly on the jet power. Although the jet power is a function of $\gamma_{j}$, it is also determined by the inject rate of rest mass. Again, we employ the 3CRR and 7CRS samples. As discussed in the last paragraph, we assume that at each redshift, 3CRR and 7CRS sources have similar age and environment density distributions as they are both complete samples. Therefore, as the 7CRS sample has a much lower flux limit, the sources in the 7CRS sample are generally less powerful than that in the 3CRR sample. In Figure \ref{l_z}, we can not see a significant difference between the size distributions from the two samples, which is in agreement with our prediction. However current observational samples are small with poor statistics. Larger samples with lower flux limit need to be considered in the future work.

\begin{figure}
\includegraphics[width=0.5\textwidth]{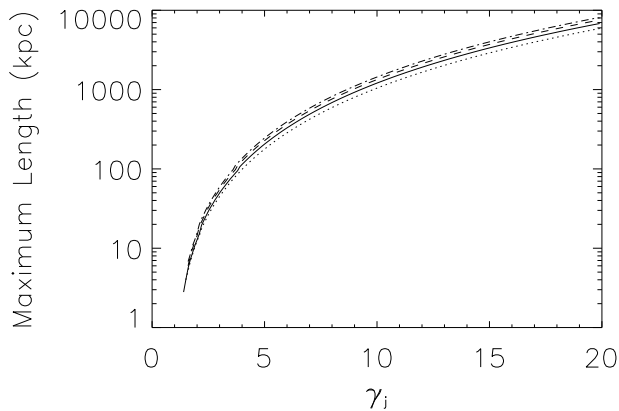}
\caption{The same diagram with Fig. \ref{gamma_l}, but with different value of $\alpha$ ($R_{T}=2$). The  dotted, solid, dashed, dash-dotted line refer to $\alpha=1.9, 1.5, 0.7, 0.0$ respectively.}
\label{different_alpha}
\end{figure}

\subsection{Dependencies on parameters}

In the previous section, we discussed about how the maximum length of a FR\,II object depends on the bulk velocity of the central jet. However, $L_{max}$ may also depend on other parameters which we set to be constants in Section 2. Some of the parameters have not been well constrained by observations, (e.g. $r_{o}$ and $\mathscr{R}_{j}$,) so we leave them as constants in this paper. However, $\alpha$ and $R_{T}$ are well constrained and studied for radio jet evolution. Meanwhile the value of $\eta$ has been discussed a lot in previous work. Therefore, we will focus on these three parameters and discuss how their values could affect the calculated $L_{max}$.

\subsubsection{The power-law index of environment density distribution, $\alpha$}

\citet{falle91} has shown that for $\alpha>2$, a jet shock could not form. X-ray observations confirm that the value of $\alpha$ should be between 0 and 2, but the values for individual objects may vary significantly. In this context, we also need to consider the core radius, $a_{0}$, inside which we take the jet to be surrounded by a constant density environment ($\alpha$ = 0). \citet{croston08} find that some FR\,I sources have fairly flat environments up to 100 kpc. Although this may not apply to FR\,IIs, it is still important to investigate how sensitive our results are to the adopted value of $\alpha$.

In order to answer this question, we calculate $L_{max}$ as a function of $\gamma_{j}$ for different values of $\alpha$ and plot the results in Figure  \ref{different_alpha}. From this, we find that $\alpha$ has only a very  small effect on the relation between $L_{max}$ and $\gamma_{j}$. If the jet is located in a flatter environment, it will have a slightly bigger maximum length and vise versa. This result is in line with our assumption that the maximum length of an FR\,II does not depend directly on its environment properties. Currently, there are not many sources with well-determined environment density profiles, but in the future it will be well-worth checking if a strong relation between $\alpha$ and average jet size exists, since this would clearly contradict with our model.

\begin{figure}
\includegraphics[width=0.5\textwidth]{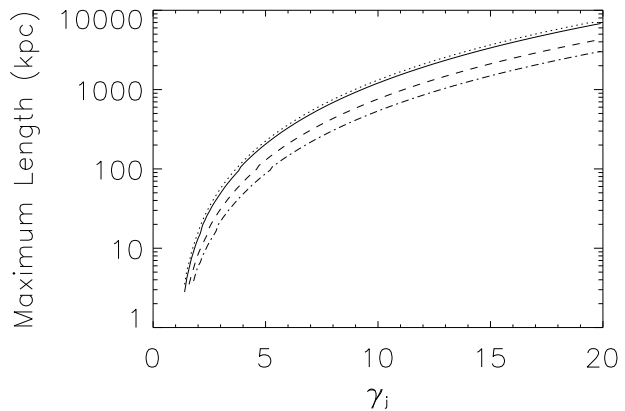}
\caption{The same diagram with Fig. \ref{gamma_l}, but with different value of $R_{T}$ ($\alpha=1.5$). The dotted, solid, dashed, dash-dotted line refer to $R_{T}=1.3, 2.0, 4.0, 6.0$ respectively.} \label{different_rt}
\end{figure}

\subsubsection{The axial ratio, $R_{T}$}

We set $R_{T}=2$ as a constant in Section 2, but it may also have different values for different objects and lead to different $L_{max}$ as well. Our evolutionary model is self-similar, so we assume $R_{T}$ is constant throughout the jet lifetime. However, this may not be true in detail, especially during the late stages of FR II evolution. Some simulations suggest that $R_{T}$ changes with time \citep{krause05}, and there is also strong observational evidence that larger sources have larger $R_{T}$ \citep{mullin08}. It is therefore also important to check how sensitive $L_{max}$ is to a variable $R_{T}$.

Figure \ref{different_rt} shows the $L_{max}-\gamma_{j}$ diagram with different values of $R_{T}$. It shows that a fatter jet should have a larger maximum length. As $R_{T}$ is used for calculating the properties inside the lobe, which directly associate with entrainment process, it is reasonable to have a bigger influence on $L_{max}$ than $\alpha$. However, the minimum and maximum suggested values of $R_{T}$ only change the final $L_{max}$ by a factor of around 3, so our assumption of constant $R_{T}$ should still be a tolerable approximation.

\subsubsection{The entrainment efficiency, $\eta$}

The entrainment efficiency is another important parameter constraining the maximum size of an FR\,II jet, but it is difficult to obtain from either theoretical analysis or observations. CR91 discussed the entrainment efficiency in the context of their model and showed that this is limited either by the ability of the environment/jet to supply material for the mixing layer, or by the maximum possible growth rate of mixing layer itself. They defined three different regimes, each associated with a particular upper limit on the entrainment efficiency. In our model, the  environment is actually the cocoon and $\rho_{c}\ll\rho_{j}$, so the appropriate regime is the \textit{environment-limited regime}. As the system is in a pressure equilibrium, we must then have a sound speed in the cocoon (environment) much larger than that in the jet. Therefore, we can only set a common upper limit of $\eta<1$. The actual value of $\eta$ for individual sources is more difficult to generalise as a number of instabilities (e.g. Kelvin-Helmholtz, current-driven and so on) together with jet properties can play important roles in setting the value of $\eta$. As a result, in our actual calculations in this paper, we have simply assumed a typical value of $\eta=0.5$. Different values of $\eta$ in the allowed range will result in a large $L_{max}$ coverage, so we do not intend to plot a $L_{max}-\gamma_{j}$ diagram here. However, it is easy to see that a lower (higher) entrainment efficiency will lead to a slower (faster) destruction of the central jet and to a larger (smaller) maximum jet size. For example, if $\eta=0.9$, a jet need to have $\gamma_{j}>12$ to reach a maximum size of 1000\,kpc, but for $\eta=0.1$, the same $L_{max}$ can be reached with only $\gamma_{j}>5$.

\subsection{Comparison with previous work}

KA97 also discussed the stability of jet based on the mixing layer model from CR91. Given that the W09 model used here is also based on CR91, it is interesting to compare our results to those obtained by KA97.

KA97 found that in an environment with $\alpha=0$, the jet can easily be destroyed and can only be stable up to 2.6 kpc. With $\gamma_{j}=2$ (KA97's default assumption), our model predicts a slightly lager $L_{max}$ of around 5 kpc for this $\alpha$. However, KA97 also argued that the $L_{max}$ will increase with $\alpha$, i.e. jets in the environment with a steeper density gradient can grow to larger sizes before becoming unstable. This contrasts with our model, in which the value of $\alpha$ plays only a minor role in setting $L_{max}$, with $\gamma_{j}$ being the key parameter instead.

There are two main differences between the models from KA97 and this paper. First, KA97 adopt \textit{mixing-layer-limited regime} of CR91 model, and obtain an upper limit of $\eta<0.26$ for the entrainment efficiency. However, as we discussed in the last section, the \textit{environment-limited regime} should be more appropriate here. Meanwhile, the original CR91 model applied by KA97 was designed for non-relativistic cases, whereas the W09 model is based on the relativistic conservation laws. In the non-relativistic case, the energy and momentum are mainly decided by the density of the lobe, which strongly depends on lobe volume and $\alpha$. However, in the relativistic case, the momentum and energy are dominated by the relativistic component, so the Lorentz factor is the key parameter.

A number of numerical simulations have studied jet instabilities in more detail taking into account more factors other than the Lorentz factor as considered in our model. For example, \citet{mizuno07} suggested that the distribution of magnetic field is crucial for determining the Kelvin-Helmholtz stabilisation. Moreover, \citet{rossi08} and \citet{meliani08} found that the environment/jet density contrast is important in determining the instability evolution and entrainment properties. However, all these simulations only represent the early stage of jet evolution. At this stage, the lobe is still small and not well established. Meanwhile, the mixing between the environment and the lobe may be significant, and the lobe density is much higher than that at the later stages of jet evolution, which are considered in this paper here. Therefore the properties of the environment play more important roles. \citet{perucho04b} performed long-term simulations for the relativistic jets, and considered a slightly overdensed environment and lobe, which is more closed to our case here. They also found that the jet Lorentz factor is an important parameter deciding the nonlinear stabilities of the jets. With the same thermodynamical properties, the jets with smaller Lorentz factors start to mix and transfer momentum to the environment at a earlier stage. However, as they tried various models with different values of thermodynamical properties, they claimed that a number of other parameters (e.g., jet-to-ambient enthalpy ratio, temperature and jet internal energy,) also affect the jet stabilities and long-term morphologies. Although in our model here, we only consider the jet Lorentz factor as the key parameter, it is worth reiterating one particular simplification we make, which is that we attribute all kinds of jet instability evolutions to the growing of the shear layer between the central jet and the lobe. In reality, the evolution and the growing speed of the shear layer must surely be a function of several other physical parameters and may therefore vary in time and between sources \citep{perucho05}. We make this simplification here purely to keep the model analytically tractable, although with sufficient data it may become possible to reconstruct the dependence of instability evolutions on various jet parameters from observations in the future.

\section{The evolution from FR\,II into FR\,I sources}
\label{model_transition}

The existence of a maximum size for FR\,II sources due to the erosion of their laminar jets raises an obvious question: what happens to an FR\,II object that reaches this limit? \citet{bick94} suggested that the transition between FR\,I/II is due to the transition from subrelativistic to relativistic flow caused by entrainment. As the death of FR\,IIs in our model here is also due to the entrainment, we argue that the FR\,IIs reaching their maximum sizes are likely to evolve into FR\,Is. In this section, we will outline a simple, but plausible scenario for the transition of a radio galaxy with an FR\,II morphology to one with an FR\,I morphology. The basic idea is sketched in Figure \ref{cartoon}.

When a stable radio outflow is born at time $t_{0}$, it exhibits an FR\,II structure with a laminar flow embedded inside a lobe and a hotspot at the end. At stage $t_{1}$, when the jet length $L_{j}$ is smaller than the maximum length $L_{max}$, the outflow grows with age, following the KA97 picture. Meanwhile, however, the central jet continuously suffers entrainment from the lobe, and the structure of the outflow can be described by our model here. The outflow evolves with an FR\,II morphology until it reaches $L_{max}$ at age $t_{max}$. At
this time, the central jet is totally eroded, and the hotspot vanishes. The detailed evolution process of the radio outflow at this stage is described in Section 2.

When the outflow evolves to an age of $t_{2}$, where $t_{2}>t_{max}$, the shear layer dominates the end region of the jet and the hotspot vanishes. A weaker shock and the lobe structure may still exist with plasma injected into the lobe after the shock from the end of the jet. The expected structure at this stage is reminiscent of a typical lobed FR\,I source. As the outflow becomes even older, the energy from the shear layer can hardly support the lobe structure or the working surface of the shock at the end of the jet, so the plasma will form a turbulent tail with the lobe disappearing either because it is refilled from the environment or because it simply runs out of energy. At the end of this evolution stage, we will observe a naked tailed jet like 3C\,31. The jet is in direct contact with the environment, and a mixing shear layer is formed. At the same time, the laminar part may shrink again as the density of the environment is higher than that of the lobe.

Please note that we are not claiming that our transition scenario here is the only way to generate FR\,Is. The precursors of FR\,Is may also include weak CSS sources \citep{pm07}, jets hitting dense environments \citep{meliani08}, and weak FR\,IIs reaching the pressure equilibrium with their environment inside the core region \citep{kb07}. These works are not in contrast with our work here as we are considering if FR\,IIs can evolve into FR\,Is at the late stage of their evolution. Our transition scenario is just complementary to all the work above, providing a new plausible way for powerful FR\,IIs to develop into FR\,Is later in their lives. More studies are still needed in order to fully understand the FR\,I/II dichotomy.

\section{Conclusion}
\label{conclusion} 

We have embedded a mixing-layer model originally developed for modelling FR\,I jets into a self-similar model for FR\,II radio lobes to study the effect of entrainment on the central jets in FR\, II objects. We find that, for reasonable parameters, the growing mixing layer between the central jet and the radio lobe could play an important role during the evolution of FR\,II objects.

The maximum length that a jet can reach is decided mainly by the bulk velocity of the particles in the central jet, $\beta_{j}$, but not directly related to the environment or the jet power. We find a maximum length of $\sim$1000 kpc for $\beta_{j}=0.997$, assuming an entrainment efficiency of $\eta=0.5$, an environment index of $\alpha=1.5$ and a lobe axial ratio of $R_{T}=2$. If the jet is located in a flatter environment with a smaller $\alpha$ or the jet is fatter with a smaller $R_{T}$, its maximum length will be larger.

We have also sketched the likely evolution of FR\,II sources after they reach their maximum size. Once the hotspots are extinguished, such sources will initially look like lobed FR\,I objects. However, ultimately their lobes will stop getting fed from the jet,  become turbulent and be refilled by the environment. At this point they will emerge as classic, 3C\,31-like FR\,I sources. This simple scenario suggests a new evolutionary connection between FR\,I and FR\,II sources and may help to shed new light on the FR\,I/II dichotomy.

The FR\,I/II transition process suggested by our model may not depend on the environment directly, but there is some observational evidence showing that FR\,Is and FR\,IIs may inhabit different in environments \citep{pp88}. Thus we cannot totally rule out the influence from the environment. There might be relations between environment properties and model dependent parameters (e.g. $\gamma_{j}$), so that the environment can affect $L_{max}$ \textit{indirectly}. This is an interesting open question for the future work once we have more observational data for the jet environment.

In closing, we stress that the picture we have developed here -- especially that of the evolution beyond $t_{max}$ -- is still basically a toy model. In the future work, we are planning to model the evolution of the jet from FR\,II to lobed FR\,I to tailed FR\,I in more detail. Our final goal is to build a unified model for all types of radio galaxies and track how they evolve and morph into each other across the P-D diagram.

\section*{Acknowledgments}
We thank Prof. Robert Laing for helpful comments. JHC acknowledges funding from the South-East Physics Network (SEPNet).
\label{lastpage}

\bibliography{wy}

\end{document}